\documentclass[a4paper]{jpconf}
\usepackage{graphicx}

\begin{document}
\title{Soft Photons from transport and hydrodynamics at FAIR energies}

\author{Andreas Grimm\textsuperscript{1,2}, Bj\o{}rn B\"auchle\textsuperscript{1}}

\address{\textsuperscript{1} Frankfurt Institute for Advanced Studies, D-60438 Frankfurt am Main, Germany}
\address{\textsuperscript{2} Institut f\"ur Theoretische Physik, Goethe-Universit\"at, D-60438 Frankfurt am Main, Germany}

\ead{grimm@fias.uni-frankfurt.de, baeuchle@th.physik.uni-frankfurt.de}

\begin{abstract}
 Direct photon spectra from uranium-uranium collisions at FAIR energies (E$_{\rm lab}$ = 35 AGeV) are calculated within the hadronic
 Ultra-relativistic Quantum Molecular Dynamics  transport model. In this 
 microscopic model, one can optionally 
 include a macroscopic intermediate hydrodynamic  phase. The hot and dense stage of the collision is then modeled by a hydrodynamical calculation.
 Photon emission from transport-hydro hybrid calculations is examined for purely hadronic matter and matter that has a cross-over phase transition and a
 critical end point to deconfined and chirally restored matter at high temperatures. We find the photon spectra in both scenarios to be  
 dominated by Bremsstrahlung. Comparing flow of photons in both cases suggests a way to distinguish 
 these two scenarios. 
\end{abstract}

\section{Motivation}

 One of the major goals in high energy nuclear physics was and is the examination of a new strongly interacting deconfined phase of matter, the 
 Quark-Gluon-Plasma (QGP) \cite{bla,bla2}, the state of matter which is believed to have filled the universe shortly after the Big Bang. Many hints for the existence of 
 this phase have 
 been found at the Relativistic Heavy Ion Collider (RHIC). The relevant observables are, e.g., jet 
 quenching and high elliptic flow \cite{bla3, bla4, bla5, bla6}. The CERN-SPS program has also found evidence for this new state of matter. One has to mention
 the enhanced K/$\pi$ 
 ratio (’horn’) and the step in the mean transverse mass excitation function for pions, kaons and protons \cite{bla7}. More recently, LHC has confirmed these results at much
 higher energies \cite{bla8}.
 The ALICE Collaboration has shown that
 the QGP remains a nearly ideal liquid even at much higher energies compared to RHIC \cite{bla9}. 

 One of the biggest challenges facing heavy-ion physicists is to get information from the hot and dense collision area, the so-called fireball. The small space-time volume
 of the fireball makes it impossible to probe it directly, so all information has to be obtained from its decay products. Single particles carry only 
 information about their last collision, which usually happens in the cold phase at the end of the collision. 
 Hence, the QGP can only be probed indirectly by most of the possible observables. 
 One of the probes suggested for searching the QGP were photons and leptons. The small interaction cross-section of electromagnetic probes leads to a mean free path 
 which is large compared 
 to the interaction zone. Thus, most photons and electrons
  leave the interaction zone undisturbed and carry information of the matter present when they were created through the entire collision.  
 Possible electromagnetic probes, which are used for measurements, are single- and dileptons and photons. 
 One further distinguishes 
 between direct photons that are created in elementary collisions, the so-called direct photons, and decay photons. This leads to the biggest challenge in measuring 
 photons in heavy-ion collisions: Most of
 the photons created in the collision are decay photons from the late stages of the reaction, namely from $\pi^{0}$ and $\eta$ decays.
 The dominant process is $\pi^{0}\rightarrow \gamma \gamma$. To gain direct 
 photon data, one has to subtract the huge background of decay photons. Some experiments have measured direct photon spectra. Explicit data points at low transverse
 momenta have been published
 by the WA98 Collaboration (CERN-SPS) \cite{bla11} and the PHENIX Collaboration (BNL RHIC) \cite{bla12,bla13}. Theoretical studies have been performed earlier, e.g., with
 the Ultra-relativistic Quantum Molecular Dynamics (UrQMD) model (see section \ref{secmodel}) by Dumitru {\it et al.}\ \cite{bla14} and
 B\"auchle {\it et al.}\ \cite{bla15} and with hadron-string dynamics (HSD) by Bratkovskaya {\it et al.}\ \cite{bla16}. Hydrodynamic approaches for direct photon calculations have been 
 used for example in: \cite{bla17, bla18, bla19, bla20, bla21, bla22}. 
 A comparison between PHENIX data \cite{bla12,bla13} and the model established in 
 \cite{bla15}  can be found in \cite{bla25}. 

 In the present work, we investigate the composition of direct photon production  and elliptic flow v$_{2}$ in U+U-collisions at E$_{\rm lab}$ = 35 AGeV using 
 the model described in \cite{bla15}.  
 We compare calculations with hadronic degrees of freedom to calculations with a first order phase transition to a QGP. 
 In Section \ref{secmodel} and \ref{secsources}, we briefly present the underlying model used for calculations and the photon sources, 
 and in Section \ref{secresult} we analyze our calculations for direct photon emission and the elliptic flow v$_{2}$. 

\section{The Model} \label{secmodel}
 This work is based on the microscopic Ultra-relativistic Quantum Molecular Dynamics  
 (UrQMD) model \cite{bla26, bla27, bla28}, which is originally a hadronic transport model with hadronic and string degrees of freedom.  In UrQMD, all non-charmed baryons and 
 mesons with masses up to \mbox{2.2 GeV} are included. 
 For resonances with high masses, the widths increase and the 
 particles overlap. Therefore, UrQMD replaces resonances with continuous string 
 excitations for higher energies. For hard scatterings, i.e., scatterings with momentum transfer higher than \mbox{Q\textsuperscript{2} = 1.5 (GeV)\textsuperscript{2}} and a
 minimal center of mass energy 
 $\sqrt{s_{\rm min}} =$ 10 GeV, PYTHIA \cite{bla29} is used. The cross-sections in UrQMD are tabulated from experimental data, 
 parametrized or, if no such data exists,
 extracted via detailed balance or the additive quark model (AQM). Microscopic models on the basis of an effective solution of the Boltzman equation can be used for 
 non-equilibrium systems and provide all information about the particle trajectory at any time of the collision. On the other hand, these kinds of models are in most cases 
 restricted to 
 binary collisions \mbox{(2 $\rightarrow$ n)}. This implies 
 large mean free paths of the particles, which may not be fulfilled  in the high-density intermediate 
 stage of the collision. 
 One possible description for the hot and dense stage of the collision is ideal hydrodynamics. High elliptic flow observed at 
 RHIC experiments seem to be consistent with ideal hydrodynamic calculations and points to the applicability of this approach \cite{bla30,bla31,bla32}.
 In UrQMD version 3.3, a hybrid approach with a micro- and macroscopic description was incorporated into the UrQMD model \cite{bla33, bla34, bla35,bla36}. 
 It is now possible 
 to switch from 
 the transport to the hydrodynamic description and back. At the beginning, the particles are propagated on straight trajectories in
 the so-called cascade mode without potentials. After the first collisions have taken place, fluidization happens and the intermediate
 hydrodynamic  stage starts. The time for this is chosen to be when the nuclei have passed through each other, i.e.,
 $t = 2R\sqrt{2m_{\rm N}/E_{\rm lab}}$ with R being the radius of the nucleus, m$_{\rm N}$ the nucleon mass and 
 E$_{\rm lab}$ the kinetic beam energy. Spectators are not fluidized. The particlization, i.e., the transition from the fluid-based to the particle-based description, starts when
 the energy density falls below a certain value in all propagated cells at the same longitudinal position. Particles are created using the Cooper-Frye formula \cite{bla37}.
 All further scatterings and decays are again performed in the microscopic UrQMD model. 
 For the 
 hydrodynamic calculations, different equations of state (EoS) with different degrees of freedom are used. The Hadron Gas EoS (HG-EoS) \cite{hg} with only hadronic degrees of 
 freedom can be applied for baseline calculations to compare with cascade calculations. The Deconfinement EoS (DE-EoS) \cite{chi} includes a first-order phase-transition to a chirally
 restored and deconfined phase of matter at high baryo-chemical potentials, a critical end point and a cross-over at low baryo-chemical potentials
 and therefore can be applied for photon emission from QGP. 

\section{Photon emission sources} \label{secsources}

 Photons are not implemented natively in the UrQMD model. Within our model, photons are calculated perturbatively without 
 changing the underlying evolution of the fireball. 
 This treatment is justified, since the electromagnetic cross-section, responsible for photon production, is negligibly small compared to the strong cross-sections 
 dealing with the evolution of the medium. 

 For the emission of photons from the hydrodynamic stage of the evolution the parametrization by Turbide {\it et al.}\ \cite{bla19} is used. The cross-sections for the 
 emission from the transport evolution description are from 
 Kapusta {\it et al.}\ \cite{bla38}. The two channels implemented in both descriptions and yielding most of the photons are namely \( \pi \pi \rightarrow \gamma \rho \) and 
 \( \pi \rho \rightarrow \gamma \pi \). The \( \pi \rho \rightarrow \gamma \pi \) rate from \cite{bla19} also includes an a$_1$-meson as an intermediate stage. 
 Emission from QGP and Bremsstrahlung processes ($\pi\pi \rightarrow \pi \pi \gamma$) \cite{brems}  are only taken into account in the hydrodynamic description. A comprehensive 
 discussion of other hadronic channels can be found in \cite{bla15}.  Photons from primordial pQCD scatterings can be neglected at FAIR energies.

\section{Results} \label{secresult}
\begin{figure}[h]
 \begin{minipage}[t]{0.48\linewidth}
  \includegraphics[width=\linewidth]{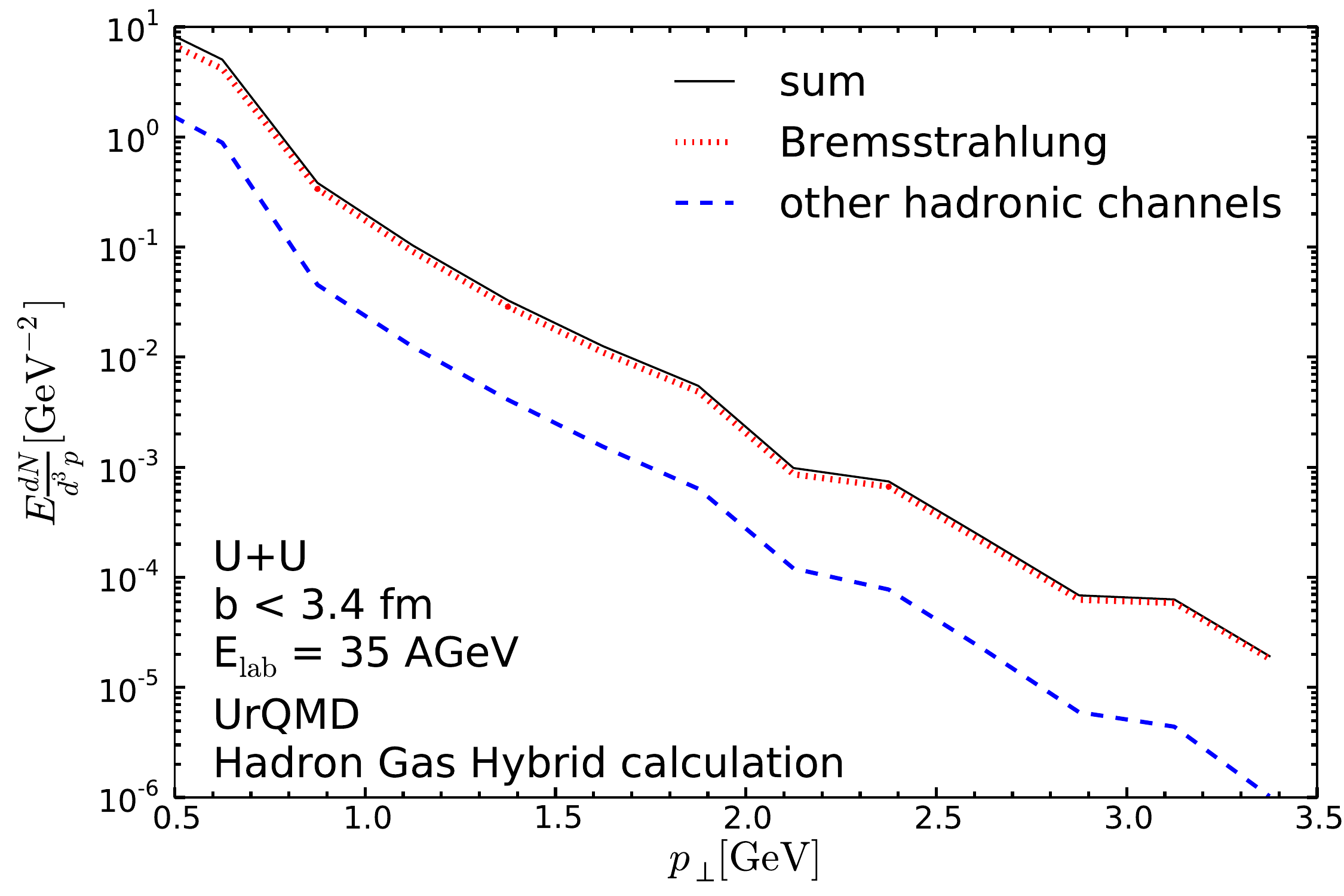} 
  \caption{\label{pic1} (Color online.) Photon emission from HG-EoS hybrid calculations of central U+U-collisions at $E_{\rm lab} = 35$~AGeV. 
           We show the total spectra (black solid line), emission from Bremsstrahlung processes (red dotted line) and other hadronic channels (blue dashed line).}
 \end{minipage}\hfill
  \begin{minipage}[t]{0.48\linewidth}
  \includegraphics[width=\linewidth]{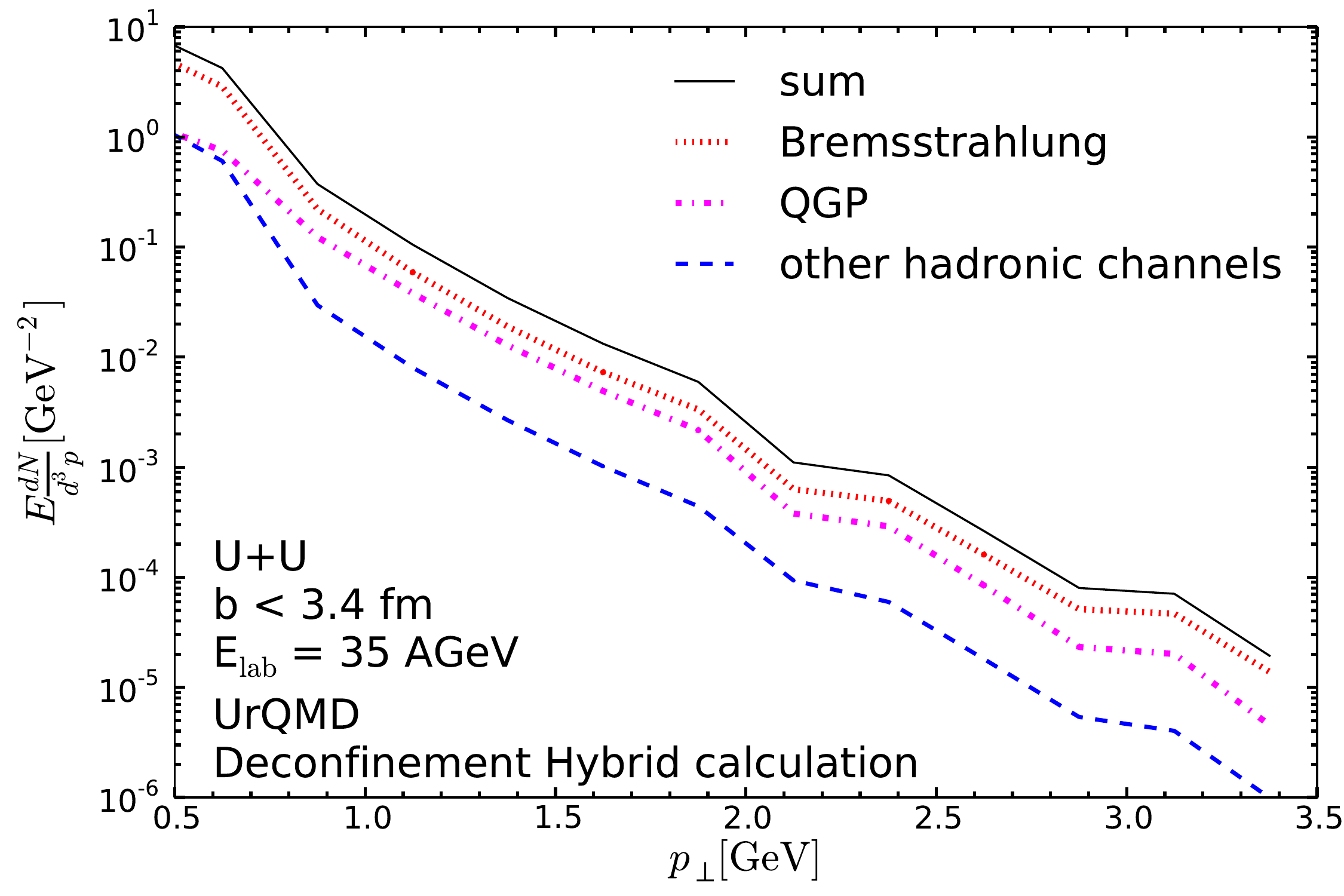} 
  \caption{\label{pic2} (Color online.) Photon emission from DE-EoS hybrid calculations of central U+U-collisions at $E_{\rm lab} = 35$~AGeV. 
           We show the total spectra (black solid line), emission from Bremsstrahlung processes (red dotted line), 
           other hadronic channels (blue dashed line) and emission from QGP (magenta dash-dotted line).}
 \end{minipage} 
\end{figure}
Figures \ref{pic1}  and \ref{pic2} show the yield of direct photons from transport and hydrodynamic calculations with the Hadron Gas EoS (HG-EoS) and the Deconfinement 
EoS (DE-EoS), respectively.
We show the spectra for photons from Bremsstrahlung (red dotted lines), QGP (magenta dash-dotted line, only Figure~\ref{pic2}), all other channels (blue dashed lines) 
and the sum of all channels (black solid lines).
Emission from Bremsstrahlung is the largest contribution to the total spectra in both cases.
In calculations with a phase transition to deconfined matter (see Figure~\ref{pic2}), the amount of photons from Bremsstrahlung is reduced, since a sizeable portion of 
the system is in the deconfined state and therefore does not emit photons from hadronic processes.
Emission from the QGP, however, makes up for the loss almost exactly, so that the total direct photon spectra from both calculations are very similar.

\begin{figure}[h]
  \begin{minipage}[t]{0.48\linewidth}
  \includegraphics[width=\linewidth]{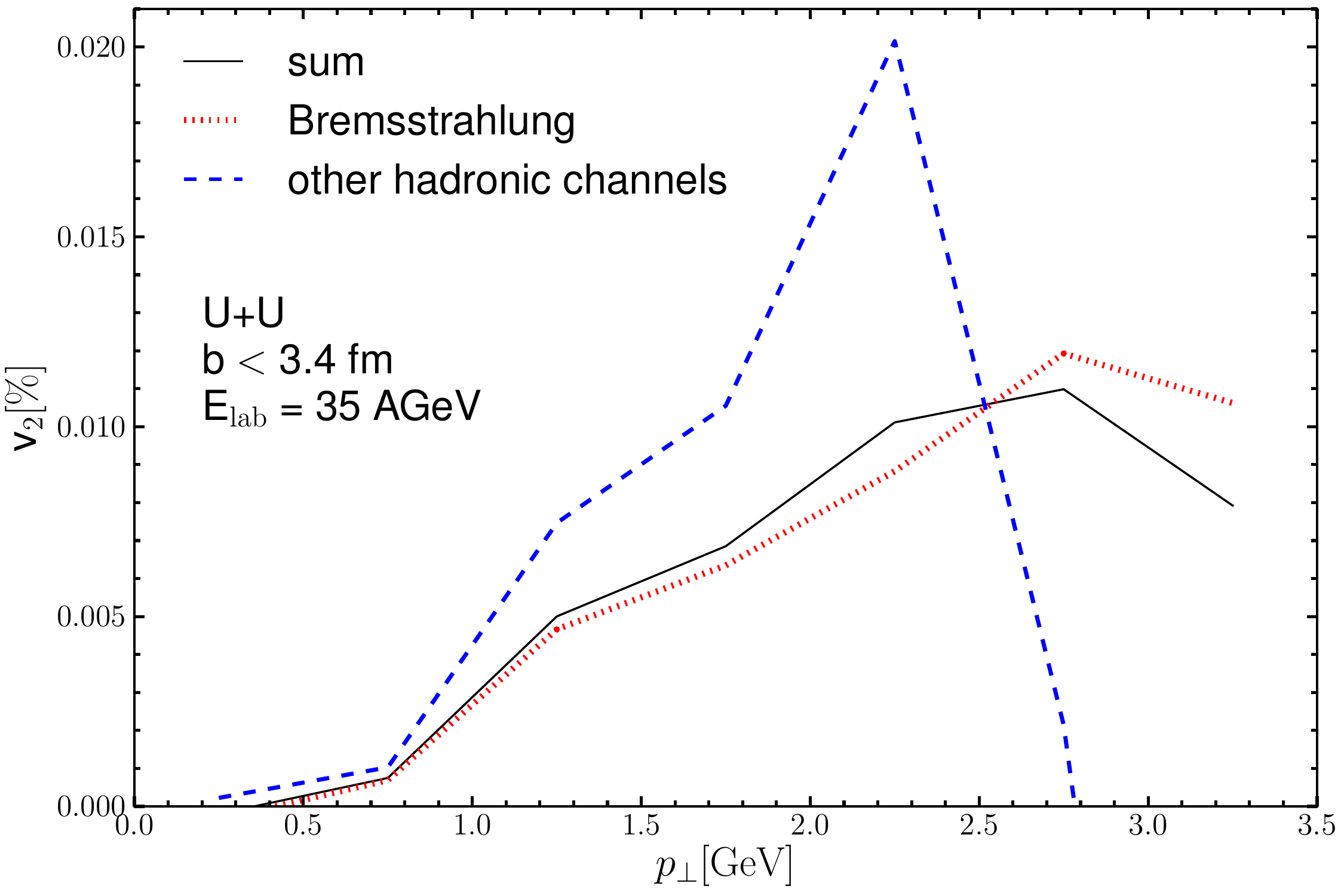} 
  \caption{\label{pic3} (Color online.) Direct photon elliptic flow v$_{2}$ for central U+U-collisions at $E_{\rm lab} = 35$~AGeV in Hadron Gas EoS Hybrid calculations. 
            We show the flow of all photons (black solid line), flow of Bremsstrahlung photons (red dotted line) and flow from other hadronic channels (blue dashed line).}
 \end{minipage}\hfill
 \begin{minipage}[t]{0.48\linewidth}
  \includegraphics[width=\linewidth]{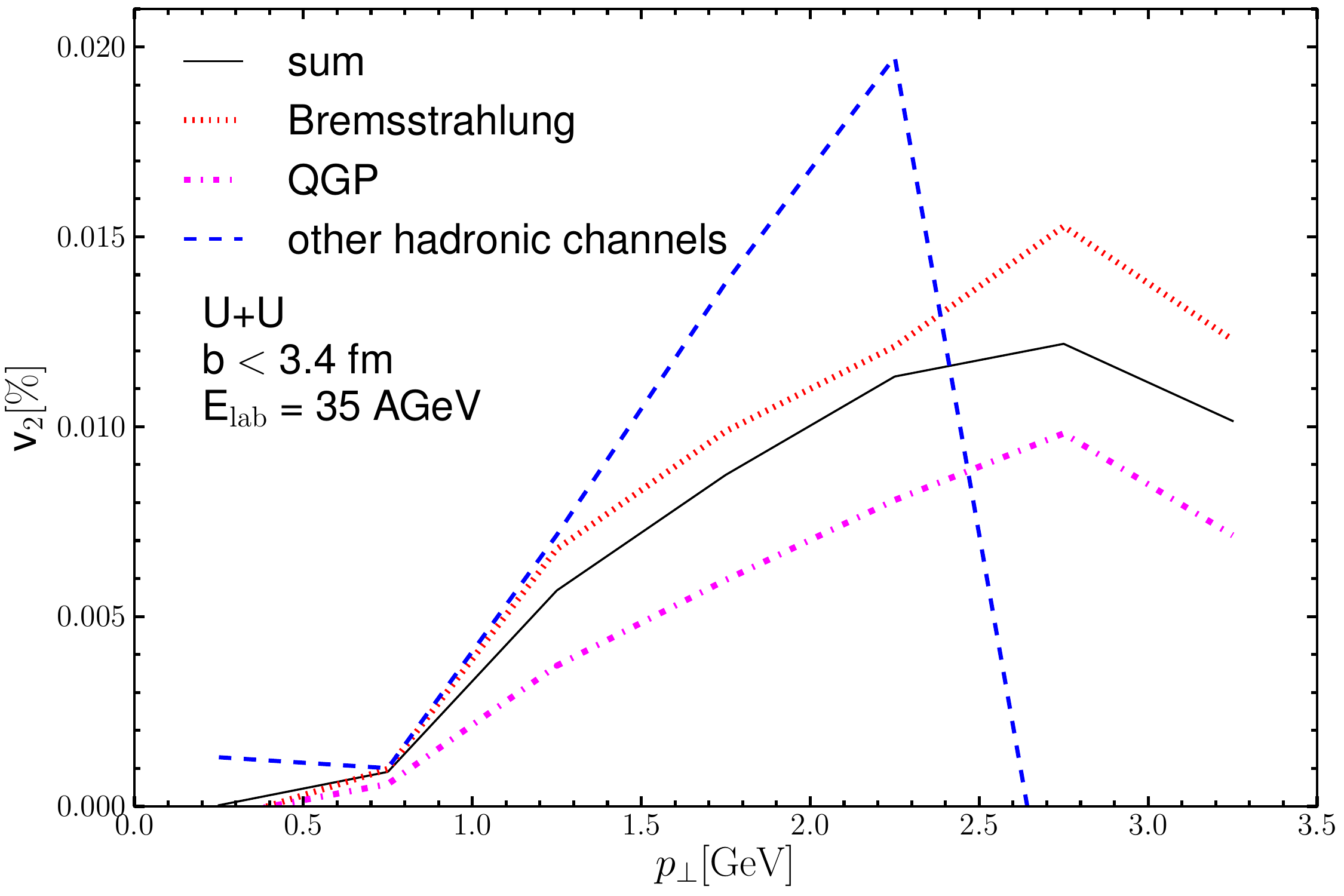} 
  \caption{\label{pic4} (Color online.) Direct photon elliptic flow v$_{2}$ for central U+U-collisions at $E_{\rm lab} = 35$~AGeV in Deconfinement EoS Hybrid calculations. 
           We show the flow of all photons (black solid line), flow of Bremsstrahlung photons (red dotted line), flow from other hadronic channels (blue dashed line) and 
           flow of photons from QGP (magenta dash-dotted line).}
 \end{minipage} 
\end{figure}
In Figures \ref{pic3} and \ref{pic4}, we show results for direct photon elliptic flow HG-EoS and DE-EoS hybrid calculations, respectively. 
We show the elliptic flow v$_{2}$ for photons from Bremsstrahlung (red dotted lines), QGP (magenta dash-dotted line, only Figure~\ref{pic2}), 
all other channels (blue dashed lines) 
and the sum of all channels (black solid lines). The elliptic flow from other hadronic channels is large compared to the flow from Bremsstrahlung and QGP in both 
scenarios.
In both scenarios, elliptic flow is dominated by the largest
contributions. The magnitude of flow from these contributions shows the
space of time in which emission from these channels happens: QGP emission
(DE-EoS calculations only) happens in the early stages of the collision,
in which the underlying hadronic medium has not yet built up strong
elliptic flow. Thus, photons emitted from this stage carry only little
elliptic flow themselves.

Photons from processes that dominate in the late part of the fireball
evolution, however, have a large hadronic flow present at their
emission time imprinted in their elliptic flow pattern. Comparing flow
results from both scenarios, we can re-assess the conclusion we drew for
direct photon spectra in Figures~\ref{pic1} and \ref{pic2}. There, we
concluded that QGP emission in the Deconfinement EoS calculations basically
substitutes Bremsstrahlung emission from Hadron Gas calculations in the
early stages. Indeed, we see a larger elliptic flow from Bremsstrahlung
in the Deconfinement EoS calculations than in the Hadron Gas EoS
calculations, pointing to a later average emission time.

Unlike in the case of the spectra, the remaining difference between both
calculations remains quite sizeable, with direct photon elliptic flow
peaking about 20~\% higher where a phase transition is present. 
\section{Summary}
We have shown calculations for direct photon emission from central
Uranium+Uranium-collisions at $E_{\rm lab} = 35$~AGeV and compared the
emission patterns from matter with a first order phase transition to
chirally restored and deconfined matter and critical end point (DE-EoS)
to that of purely hadronic matter without phase transition. We find the
direct photon spectra to be essentially unchanged, while we predict
direct photon elliptic flow to be sensitive to the choice of matter.
High-quality experiments, as they are being built at the FAIR facility,
will be able to distinguish both scenarios.
    
\section{Acknowledgments} 
This work has been supported by the Frankfurt Center for Scientific Computing (CSC), the GSI and the
BMBF. We would also like to thank for the support  
of the Hessian LOEWE initiative
through the Helmholtz International Center for FAIR.

\end{document}